\documentclass{article}
\usepackage[utf8]{inputenc}
\usepackage[draft]{changes}
\usepackage{amsmath}
\usepackage{amsfonts} 
\usepackage{extarrows}
\usepackage{mathtools}
\usepackage{amsthm}
\usepackage{lipsum}

\usepackage[english]{babel}

\usepackage{subcaption}
\usepackage{wrapfig}
\usepackage[hidelinks]{hyperref}
\usepackage{url}
\usepackage{graphicx}
\usepackage{xcolor}

\theoremstyle{remark}

\providecommand{\keywords}[1]
{
  \small	
  \textbf{\textit{Keywords---}} #1
}

\title{Possible counter-intuitive impact of local vaccine mandates for vaccine-preventable infectious diseases}
\author{Maddalena Don{\`a} and Pieter Trapman\\
{\normalsize \it
Faculty of Science and Engineering, University of Groningen, NL}
}
\date{\today}

\begin{document}

\maketitle

\begin{abstract}
We model the impact of local vaccine mandates on the spread of  vaccine-preventable infectious diseases, which in the absence of vaccines will mainly affect children. Examples of such diseases are measles, rubella, mumps and pertussis. 
To model the spread of the pathogen, we use a stochastic SIR (Susceptible, Infectious, Recovered) model with two levels of mixing in a closed population, often referred to as the household model. In this model individuals make local contacts within a specific small subgroup of the population (e.g.\ within a household or a school class), while they also make global contacts with random people in the population at a much lower rate than the rate of local contacts.

We consider what happens if schools are given freedom to impose vaccine mandates on all of their pupils, except for the pupils that are exempt from vaccination because of medical reasons.
We investigate how such a mandate affects the probability of an outbreak of a disease and the probability that a pupil that is medically exempt from vaccination, gets infected during an outbreak. 
We show that if the population vaccine coverage is close to the herd-immunity level then both probabilities may increase if local vaccine mandates are implemented. This is caused by unvaccinated pupils moving to schools without mandates.
\end{abstract} \hspace{10pt}


\section{Introduction}

\subsection{Background}

Measles is a viral disease that yearly kills over a hundred thousand people worldwide \cite{W.H.O.measles}. It is one of the most contagious diseases known, with a basic reproduction number $R_0$ estimated to be between 12 and 18 \cite{Thom23} (see Section  \ref{Sec:Model}  and \cite{diekmann2013mathematical} for a definition of $R_0$). Because measles is extremely contagious and immunity after infection is typically life-long, in the absence of vaccination, most people will be infected before reaching adulthood, hence measles is often called a childhood disease. 

In some countries measles and other vaccine-preventable infectious diseases, such as rubella, mumps and pertussis, are still endemic. That is, they are always present and cause regular peaks of prevalence. In many other countries, measles is largely absent because of successful vaccination campaigns \cite{W.H.O.measles,W.H.O.vaccine}, which caused \textit{herd-immunity} in the population \cite[p69]{diekmann2013mathematical}. We focus on those countries. In several of the countries in which there have not been considerable outbreaks of measles for decades, the vaccine coverage has been declining, and there is fear that the ``pool of susceptibles'' will grow big enough to loose herd-immunity and allow for new waves of infection. Governments and public health authorities try to tackle this by campaigns to promote vaccine uptake and sometimes by legislation which makes some vaccines compulsory for children, e.g.\ in Italy \cite{MinisterodSalute}, or considering giving schools/daycare centers the freedom to use a vaccine mandate, as was considered in the Netherlands \cite{Eerstekamer}.

In this paper we consider the case in which vaccination is not mandatory for everybody and the decision of requiring the vaccine is left to the schools as was proposed in the Netherlands. In this situation, children who cannot get a vaccine for medical reasons can  freely choose where to attend school, whether it has a mandate or not.
We refer to those children as \textit{medically exempt} children. 
So, in this situation unvaccinated children that are not medically exempt all go to schools without a mandate, and in those schools the fraction of unvaccinated children may be higher than if vaccine mandates are not allowed, which (as we show) may lead to higher risk of outbreaks of the vaccine-preventable diseases.  
  
It is likely that medically exempt children go to schools that require vaccination, in order to obtain a safer environment. 
We are interested in the question whether and under which conditions local vaccine mandates will lead to increased risk of infection for unvaccinated children that are medically exempt.

\subsection{Model}\label{Sec:Model}

In this paper we use a mathematical model to analyze the spread of measles among elementary school-aged children, where we take into account that most spread of the pathogen takes place within classes. Although we take measles as the pathogen of interest, the analysis applies equally to other highly infectious vaccine preventable diseases. 
We use a stochastic SIR (Susceptible, Infectious, Recovered) epidemic model in which the relevant population is closed and partitioned in school classes.
So, the total number $N$ of children is fixed and we do not consider migration, births or deaths and we assume that the time scale of an outbreak is such that we may ignore children moving to different classes. We define $n_c$ as the number of children per class, which we, for convenience of exposition, consider fixed, and $m_c$ as the number of classes. So $N=n_c m_c$. In Appendix \ref{appendix:random class sizes} we show that the mathematical analysis is easily expanded to allow for variation in the class sizes. We remark that strictly speaking we analyse our model for a sequence of populations, where $N \to \infty$.

In SIR models each individual can be \textit{susceptible} to the disease, \textit{infectious} or \textit{recovered} and immune. A contact trough which the virus is spread is called an \textit{infectious contact}. An infectious individual remains so for a random length of time that is called the \textit{infectious period}, during which they are able to infect other individuals in the population. In our analysis we only focus on the probability that an introduction of a virus may lead to a large outbreak and on the probability that a given child gets infected. We do not consider the dynamics of such an outbreak in real time. Because of this, possible exposed/latent periods, between a child getting infected and becoming infectious \cite{andersson2012stochastic} have no impact on our quantities of interest.

In many epidemic models it is assumed that the population is homogeneous, which means that the individuals mix uniformly. However, in real life this is rarely the case. Here we introduce heterogeneity by allowing individuals to have local (within class) and global contacts (between classes) with different probabilities.
We assume that, conditioned on an individual being infected, whether or not infectious contacts are made with different individuals are independent. So, we implicitly assume that the infectious period is not random. Obviously, the infectious period for measles is random. However, the variance of time between the start of the infectious period and the onset of rash (which is close to the period that a child stays socially active and is likely to infect other children) is very small \cite{Hubs22}. This makes that we can simplify the analysis and assume that contacts between individuals occur independently \cite{Kuul82}.
In our model an infectious individual makes an infectious contact with each classmate with probability $p_L$ and with other individuals in the population with probability $p_G$. This model is also referred to as the \textit{household model}, see e.g.~\cite{10.2307/2245132} or \cite[Section 6.3]{andersson2012stochastic}, since people are divided into small groups (such as workplaces, households or classrooms) where transmission of the virus is more frequent. 
The global transmission probability is typically much smaller than the local probability, and in our case we assume that $p_G$ is asymptotically inversely proportional with the population size.  That is, if we consider a sequence of epidemic models indexed by the population size $N$, then  there exists a constant $\lambda_G \in (0, \infty)$ such that $\lim_{N \rightarrow \infty} N p_G= \lambda_G$.

The basic reproduction number $R_0$ is the expected number of infections caused by a ``typical'' infected individual infected during the initial stages of an epidemic in a population that is mostly susceptible \cite{diekmann2013mathematical}. (So,  $R_0$ refers to a situation in which there is no vaccination.) In homogeneously mixing populations defining $R_0$ is straightforward. Although there is some issue with deciding what a typical infected individual is, it is also possible to define $R_0$ and other reproduction numbers for household models \cite{PELLIS201285,BALL2016108}. In Section \ref{Sec:Analysis}  we deduce the model parameter $\lambda_G$ from estimates of $R_0$ from literature.

For measles, the basic reproduction number $R_0$ is estimated to be between 12 and 18 \cite{Thom23,Hubs22}, which makes the virus one of the most contagious viruses in the world.  We simplify our model by using the  approximation $p_L = 1$, which means that if one
of the unvaccinated children in a class gets the disease, then they will infect all of their unvaccinated classmates.
Setting $p_L<1$, would complicate the exposition considerably, while the qualitative behavior of the model stays similar, as long as $p_L$ is large enough (say larger that $4$ divided by the class size, when the probability that entire class gets infected is roughly $98\%$), see Appendix \ref{appendix:pL<1}.

Because $p_L=1$ a child and their non-vaccinated classmates together can be seen as a ``super individual'',
of which the susceptibility and infectivity is proportional to the number of unvaccinated children in the class.
We assume that the latent period of measles is relatively long, so that we can safely assume that all secondary infections in a class are directly caused by the initially infected child in the class, and not by another secondary infected in the population. 
Lastly,  we define a time unit as the fixed infectious period \cite[p4]{andersson2012stochastic}.

To analyze the impact of vaccination (or the lack thereof), 
we assume that the vaccine against measles is perfect, i.e., once vaccinated a person can neither get infected nor infect other people.
So, vaccinated people are effectively omitted from the model. 
A fraction $v$ of the population is vaccinated; we call $N_u(v)=(1-v) N$ the number of unvaccinated children, which are the initially susceptibles in our model. Another way to introduce vaccines is to say that each child is independently vaccinated with probability $v$.
The two assumptions lead to slightly different models, but since they are asymptotically equivalent, we use one or the other depending on what is more convenient in the situation. For example, when it is useful to think about $N_u$ as a fixed number, we will use the first assumption; if we want the number of initially susceptible children in different classes to be independent, then the second model is better.

In order to protect children at a school, the school may (if allowed by law) institute a local vaccine mandate, which require all pupils that are not medically exempt to be vaccinated. In our model we denote the fraction of classes with such a mandate by $\pi$.

Further recall that we require $\pi \leq v$ for reasons of consistency.

\section{Analysis}\label{Sec:Analysis}

To evaluate the global infection rate $\lambda_G$ for given $R_0$, we consider a situation with no vaccines nor mandates, i.e., both $v$ and $\pi$, the fraction of classes with a mandate, are set to zero. Recall that classes are all of fixed size $n_c$. We distinguish between children that are infected through a global contact and children that are infected within a class through a local contact. The first group of children are the \textit{primary cases} or \textit{type-one individuals}, that are the first individuals of their classes to get infected; each primary case will then infect all the $n_c-1$ other (unvaccinated) children in their class, which we refer to as \textit{secondary cases} or \textit{type-two individuals}. We assume that both types make global contacts (and cause primary cases) at rate $\lambda_G$, but only primary cases cause secondary cases.
For this household model, it is known that the epidemic threshold parameter $R_0$ is the dominant eigenvalue of the \textit{next generation matrix}
\begin{equation}
\label{matrix}
A= \begin{bmatrix}
    \lambda_G & n_c-1 \\
    \lambda_G & 0
\end{bmatrix}
\end{equation}
where element $A_{i,j}$ is given by the mean number of type-$j$ individuals that a type-$i$ individual infects during their infectious period (\cite{BECKER1995207,10.2307/2245132}).
So $R_0$ 
solves
$$
-(\lambda_G -R_0)R_0 - \lambda_G (n_c-1)=0.
$$
So,
\begin{equation}\label{R0form}
\lambda_G= \frac{R_0 ^2}{R_0 +n_c-1}
\end{equation}

If $R_0=15$ and $n_c=25$, this leads to $
\lambda_G \approx 5.8.
$

Note that in \cite{BECKER1995207}, the dimension of the next generation matrix is $n_H \times n_H$, where $n_H$ is the maximal household size. In our case it is enough to consider a $2 \times 2$ matrix because we only need to divide individuals into \textit{primary} and \textit{secondary} cases. In Appendix \ref{appendix:pL<1}, when dealing with $p_L<1$, we need higher dimensional matrices.

The dominant eigenvalue of the matrix $A$ is sometimes called the individual reproduction number and is denoted by $R_I$ \cite{BALL2016108}. Assuming $p_L=1$,  the reproduction numbers  $R_0$ and $R_I$ are the same because all secondary cases are infected by primary ones only.

For the analysis of the impact of vaccination in the population, we call the fraction of the population that is vaccinated $v$, and we define $\pi$ as the fraction of classes with a vaccine mandate.
As mentioned before, there are in total $m_c$ classes, each with $n_c$ students. So, $N=m_cn_c$. The number of classes without mandate is $m_c(1-\pi)$ so the maximum possible number of unvaccinated children is $m_c n_c (1-\pi)=N(1-\pi)$. Therefore, the number of unvaccinated children $N(1-v)$ has to be less or equal than $N(1-\pi)$. So, for the model to be consistent we require $\pi \leq v$.

\begin{figure}
\includegraphics[width=8cm]{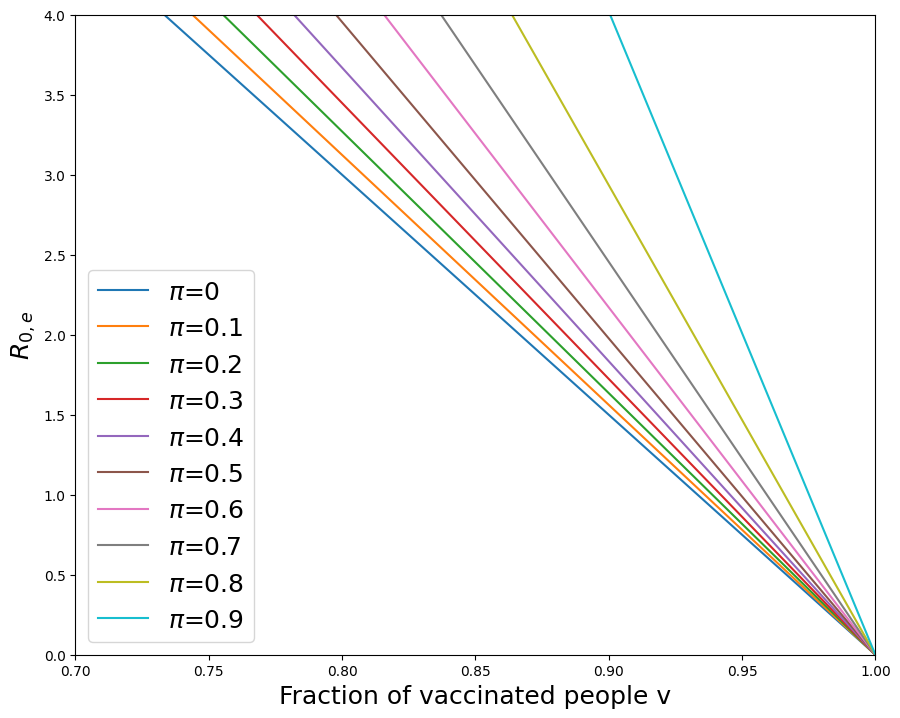}
\centering
\caption{$R_{0,e}$ as a function of $v$, if both vaccines and mandates are introduced.}
\label{B}
\end{figure}

We are interested in how the reproduction number $R_{0,e} = R_{0,e}(v, \pi)$  (where the $e$ stands for effective) changes once the vaccine is introduced in the population and depends on $v$ and $\pi$. To analyse this, we compute the dominant eigenvalue of the next generation matrix
$$
A' =A'(v,\pi) = \begin{bmatrix}
    \lambda_G(1-v) & (n_c-1)\frac{(1-v)}{(1-\pi)} \\
    \lambda_G(1-v) & 0
\end{bmatrix}.
$$
The results are depicted in Figure \ref{B}.

In a class without a vaccine mandate, the distribution of the number $X$ of unvaccinated children is $Bin(n_c, \frac{1-v}{1-\pi})$ and we write $\pi_k=P(X=k)$. In what follows we use
$$u:=\frac{1-v}{1-\pi}.$$
The expected number of unvaccinated children per class is given by $\mu_{c}=n_c u = n_c \frac{1-v}{1-\pi}$.
We observe that the probability that a child, chosen uniformly at random among the unvaccinated children in the population, belongs to a class with $k$ unvaccinated children is given by the size-biased distribution, determined through
\begin{equation}\label{binomial}
\Bar{\pi}_k=\frac{k \pi_k}{E[X]}= \frac{k \binom{n_c}{k} (u)^k (1-u)^{n_c - k}}{n_c u}
=  \binom{n_c -1}{k-1} u^{k-1} (1-u)^{n_c-1-k+1}.
\end{equation}
We notice that these are the probability masses of $Y+1$, where $Y =Y(v)$ is a random variable with a binomial distribution with parameters $n_c -1$ and  $u$.

\begin{figure}[h]

\begin{subfigure}{0.49\textwidth}
\includegraphics[width=0.9\linewidth, height=6cm]{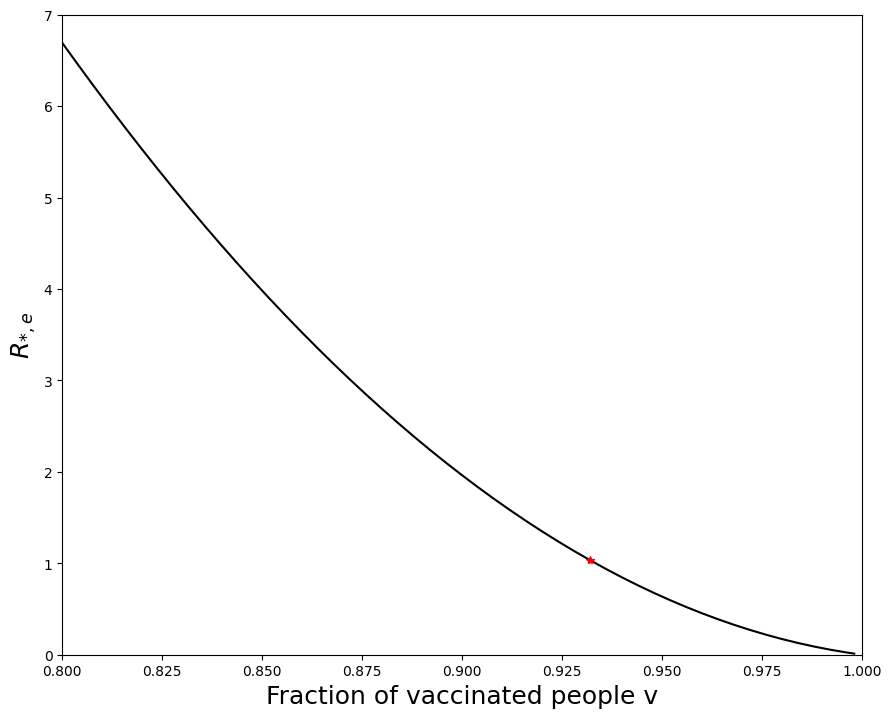} 
\end{subfigure}
\begin{subfigure}{0.49\textwidth}
\includegraphics[width=0.9\linewidth, height=6cm]{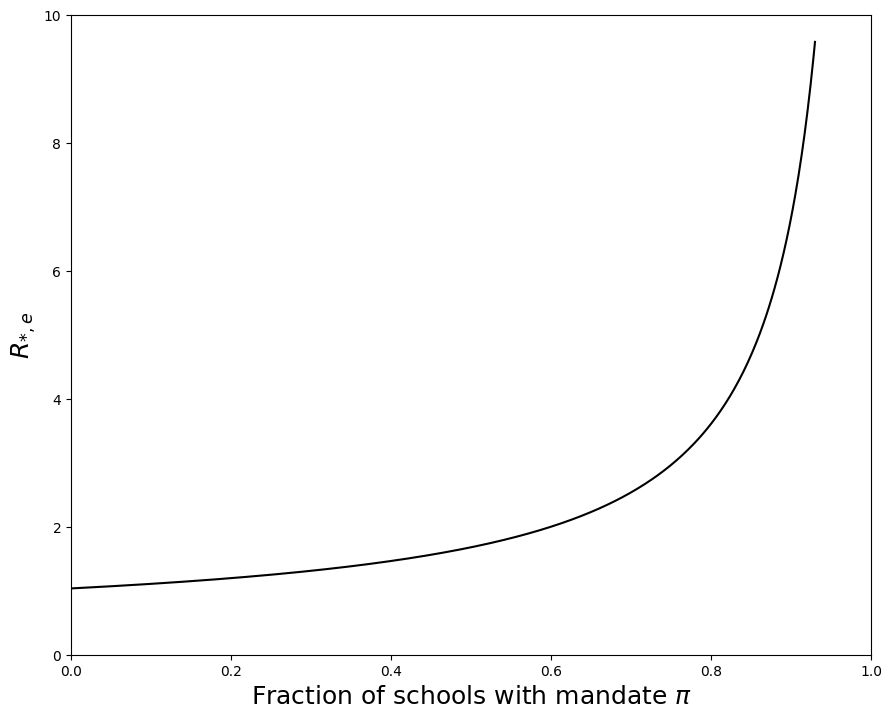}
\end{subfigure}
\caption{Left: plot of $R_{*,e}$ as a function of $v \in [0.8,1]$ when $R_0=15$, $n_c=25$ and $\pi=0$. The red asterisk denotes when $R_{*,e}=1$ and we call $v_{cr}$ the correspondent vaccination coverage.\\ Right: plot of $R_{*,e}$ as a function of $\pi$ when $v=v_{cr}$, again for $R_0=15$ and $n_c=25$. }
\label{1}
\end{figure}

Using  $\lambda_G$ from \eqref{R0form}, we compute $R_{*,e}=R_{*,e}(v,\pi)$ which is the \textit{household reproduction number} after vaccination for the household model  \cite{10.2307/2245132,BALL2016108}, and represents the mean number of clumps (or in our case, classes) that are infected by a typical infectious clump/class. It is defined to be $R_{*,e}=\lambda_{G,v} \mu$, where $\lambda_{G,v}$ is the expected number of global infections caused by a typical infected individual, and $\mu=\mu(v)$ is the mean number of infected children in an infected class. $\lambda_{G,v}$ is given by $\lambda_G (1-v)$, while
$$
\mu=\mathbb{E}[\Bar{X}]=\sum_{k=1}^{n_c} k \Bar{\pi}_k= 1+ (n_c-1)u.
$$
The latter follows from the size biased distribution in \eqref{binomial}.
Substituting this in the expression for $R_{*,e}$, we obtain
\begin{equation}
\label{R*}
R_{*,e}= [1+(n_c -1)u](1-v)\lambda_G.
\end{equation}
The household reproduction number $R_{*,e}$, as well as most other reproduction numbers defined in literature, including $R_{0,e}$, is useful because an epidemic goes extinct with probability 1 if and only if $R_{*,e} \leq 1$ \cite{BALL2016108}. The minimal value of $v$ for which $R_{*,e} \leq 1$, is \textit{the critical vaccination coverage} $v_{cr}$. 
$$
R_{*,e}>1 \Longleftrightarrow 
 \frac{(1-v)^2}{(1-\pi)}\lambda_G (n_c -1) > 1 - (1-v)\lambda_G.$$
 Since $n_c$ represents the number of children in a class, the left hand side is always non-negative. So, if the expected number of contacts with unvaccinated children (i.e., $(1-v)\lambda_G$) is greater than $1$, i.e. $ 1-v > \frac{1}{\lambda_G}$, 
  then the epidemic is above threshold. Assume instead that both terms are positive; then this inequality gives us an upper bound for the fraction $1-\pi$ of schools without mandate:
 $$
 1-\pi < \frac{(1-v)^2 \lambda_G (n_c-1)}{1-(1-v)\lambda_G}.
 $$
 So, if the number of schools without mandate is small enough and the unvaccinated children are concentrated in those schools, the value of $R_{*,e}$ is above 1 (see Figure \ref{1}).

\subsection{Branching process approximation}

\subsubsection{The branching process}

Assume that the epidemic starts with one randomly chosen unvaccinated child in the population and say that the class of the child has $\ell$ non-vaccinated children. Then the probability that a class with $k$ unvaccinated children gets infected by the class of the initially infected child is given by the following expression
\begin{equation}\label{beforeapprox}
\frac{\pi_k[1-(1-p_G)^{k \ell}]}{\sum_ {k'=0}^{n_c}\pi_k'[1-(1-p_G)^{k' \ell}]}.
\end{equation}
When $p_G <<1$ (as we assume), the expression $1-(1-p_G)^{k \ell}$ is well approximated by $k\ell p_G$, and we obtain that  \eqref{beforeapprox} is approximated by
$$
 \frac{\pi_k p_G k \ell}{\sum_{k'=0}^{n_c} \pi_k' p_G k' \ell}=\frac{\pi_k k }{\sum_{k'=0}^{n_c} \pi_k'  k' }=\Bar{\pi}_k.
$$

If the number of classes is large,  also the population size $N$ and the number of unvaccinated children $N_u$ are large. So, we can use a branching process approximation to describe the initial stages of the epidemic: in those early stages infectious individuals contact only individuals that -- with large probability -- are still susceptible, so the number of new infections caused by households are asymptotically independent and  global contacts lead to a tree-structured epidemic generated graph, which has the law of a Galton-Watson tree with offspring distribution described by $\bar{\pi}_k$ ($k \in \mathbb{N}_0$) \cite{Brit19}. 

More formally, by a birthday problem approach \cite[p.24]{Grim01}, we can argue that the probability that the first $k$ global contacts, $ k=o(\sqrt{N_u})$, are all with different households.
Furthermore, it is well known (e.g.\ \cite[Chapter 3]{andersson2012stochastic}) that if we denote the number of infected children in an SIR epidemic by $\lvert \mathcal{E}_{N_u}\rvert$ and the size of the  approximating branching process (i.e.\ a Galton-Watson process with  offspring distribution described by $\bar{\pi}_k$ ($k \in \mathbb{N}_0$)) by $\lvert \mathcal{B}\rvert$, then as $N_u \to \infty$
$$
\mathbb{P}(\lvert \mathcal{E}_{N_u}\rvert > \log N_u) \to \mathbb{P}(\lvert \mathcal{B}\rvert= \infty).
$$
If $\lvert \mathcal{E}_{N_u}\rvert > \log _uN$ we talk about a major outbreak of the epidemic.

\subsubsection{The probability of avoiding infection for a medically exempt child}

If the first infectious child belongs to a class of size $k$ (i.e.\ a class with $k$ unvaccinated children), then the probability that the process dies out is equal to the probability that the offspring of each student in the first infectious class dies out. Since the infection rate for global contacts between unvaccinated is $\lambda_G(1-v)$, a class of size $k$ will make global contacts at a rate $\lambda_G(1-v) k$. This leads to the following expressions for $\Bar{q}$, the extinction probability of the approximating branching process for an epidemic initiated by a uniformly chosen unvaccinated individual from the population.
\begin{equation}
\label{qbar}
\begin{aligned}
\Bar{q} & =\sum_{k = 1}^{n_c} \mathbb{P}(\Bar{X}=k) \sum_{\ell=0}^\infty \frac{[\lambda_G(1-v) k]^\ell}{\ell!}e^{-\lambda_G(1-v) k}\Bar{q}^l\\
 & =\sum_{k = 1}^{n_c} \mathbb{P}(\Bar{X}=k)e^{-\lambda_G(1-v) k(1-\Bar{q})}\\
& = \sum_{j = 0}^{n_c-1} \mathbb{P}(\Bar{X}-1=j)e^{-\lambda_G(1-v) (j+1)(1-\Bar{q})}\\
& =\sum_{j = 0}^{n_c-1}  \binom{n_c -1}{j}u^j (1-u)^{n_c-1-j}e^{-\lambda_G(1-v) (j+1)(1-\Bar{q})}\\
& = (1-u+ue^{-\lambda_G(1-v)(1-\Bar{q})})^{n_c-1} e^{-\lambda_G(1-v)(1-\Bar{q})}
\end{aligned}
\end{equation}
which means that $\Bar{q}$ is a solution of the fixed point equation
$$
\Bar{q}= (1-u+ue^{-\lambda_G(1-v)(1-\Bar{q})})^{n_c-1} e^{-\lambda_G(1-v)(1-\Bar{q})}.
$$
From the theory of branching processes \cite{Jage75}, we know that $\Bar{q}$ is the smallest solution of this equation, which is strictly less than 1 if and only if $R_{*,e}>1$. 

By assumption a medically exempt child is enrolled in a school that has a vaccine mandate. This means that they belong to a class of effective size 1.
By the epidemic generated graph representation of an epidemic \cite{Brit19}, we know that $p_{me}$, the probability that  a medically exempt child avoids being infected if a large outbreak occurs, is equal to the probability that a large outbreak occurs if the outbreak starts with a single infected child in a class of effective size 1. This result can also be obtained by using susceptibility sets \cite{ball_lyne_2001} and see that the so-called forward and backward branching process have the same offspring distribution.
Using arguments as in \eqref{qbar}, we obtain that 
\begin{equation}\label{xdef}
p_{me}=p_{me}(v,\pi)= e^{-\lambda_G (1-v)(1-\Bar{q})}.
\end{equation}
If there is no vaccine mandate, a medically exempt child is just like any other unvaccinated child and the probability of avoiding infection is $\Bar{q}$ for the case that $\pi =0$. We denote this probability by $p_{wm}=p_{wm}(v)$, where \textit{wm} stands for ``without mandate''.

\section{Results}

Considering the entire population, we may expect that mandates make the global situation worse, because unvaccinated children have on average more other unvaccinated children in their class and children in the same class contact each other at a higher rate than other pairs of children. 
But in this situation a medically exempt child can only be infected by global contacts. In the case without mandate, even if the global situation is more controlled, such a child could be enrolled in a class with other unvaccinated children. And the probability $p_{me}$ for a given medically exempt child to avoid infection in a situation with vaccine mandates may be larger than that probability in a situation without mandates $p_{wm}$. In this section we analyse the model to justify this intuition.

\begin{figure}
\includegraphics[width=8cm]{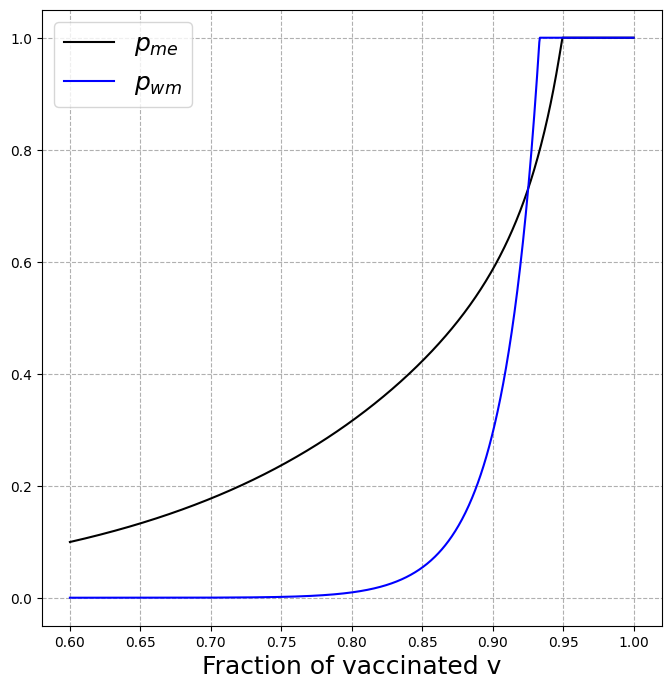}
\centering
 \caption{Plot of the probability $p_{me}$ that a medically exempt child avoids infection as a function of $v$ with $\pi$ set to $0.5$, and of the probability $p_{wm}$ that a random unvaccinated child avoids infection in the case with no mandates ($\pi=0$). We see that the two lines cross at $v_s \approx 0.925$. The minimal value of $v$ for which  $p_{wm}=1$ is at $v \approx 0.933$, while the minimal value of $v$ for which  $p_{me}=1$ is at $v \approx 0.95$.}
\label{C}
\end{figure}

In Figure \ref{C} we consider the case where $\pi=1/2$ and  $v$ is varying. Here $p_{me}$ is higher than $p_{wm}$ up to $v \leq v_s \approx 93 \%$, and we have the opposite inequality for $v >v_s$, where the $s$ stands for switching level. The graph tells us that if $v$ is low, a medically exempt child is profiting from vaccine mandates at schools, since they will for sure be enrolled in a class with no other unvaccinated children. But when the vaccine percentage is sufficiently high the probability of having only one unvaccinated child per class is already high and mandates worsen the global situation for everyone. Note that $p_{wm}$ reaches $1$ at a lower $v$ value than $p_{me}$. This confirms the intuition that without mandates a lower vaccine coverage  is sufficient to guarantee herd-immunity.

\begin{figure}
\includegraphics[width=.45\linewidth]{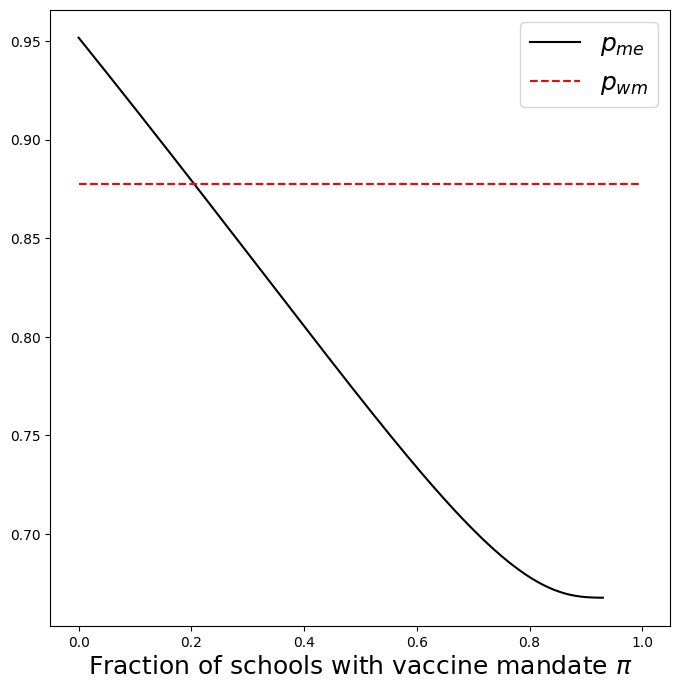} \hspace{1cm}
\includegraphics[width=.45\linewidth]{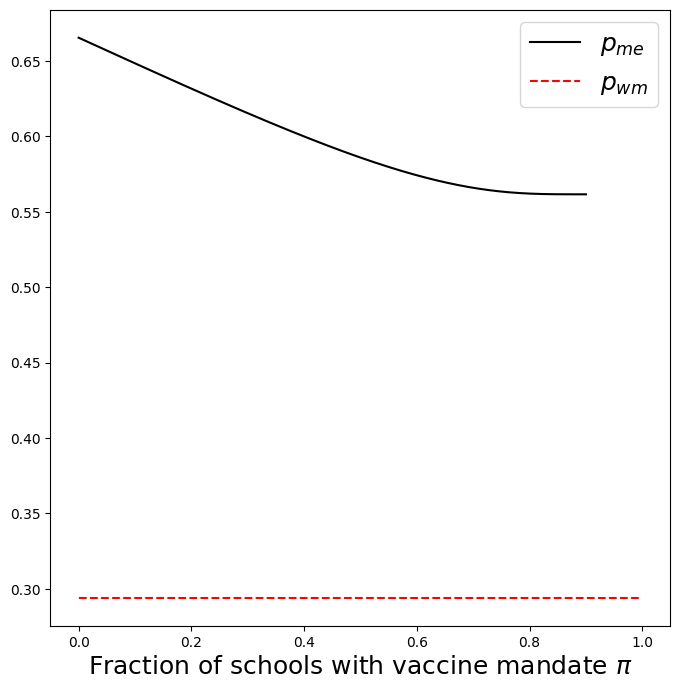}
 \caption{Plot of $p_{me}$, the probability that a medically exempt child avoids infection as a function of $\pi$, when $v=0.93$ (left) and when $v=0.90$ (right). We compare it with the probability $p_{wm}$ that a random unvaccinated child avoids infection in the case with no mandates ($\pi=0$). }
\label{D}
\end{figure}

In Figure \ref{D}, we again consider $p_{me}$, the probability that a medically exempt individual avoids infection, but now as a function of $\pi$, when $v=0.93$ and $v=0.90$. For $v=0.93$, we can see that, for low values of $\pi$ (lower than approximately $0.20$), the medically exempt are more protected in the model with mandates. But if $\pi$ increases, then the mandates are actually disadvantageous to the medically exempt children, even if they are the only unvaccinated children in their class.
If $v=0.90$, already without any vaccine mandates, the global spread of the disease is so easy, that the global spread does not increase much by introducing vaccine mandates, while the relative protection of having no other unvaccinated children in the class is considerable and vaccine mandates are always beneficial for medically exempt children.  

\begin{figure}
\includegraphics[width=.5 \linewidth]{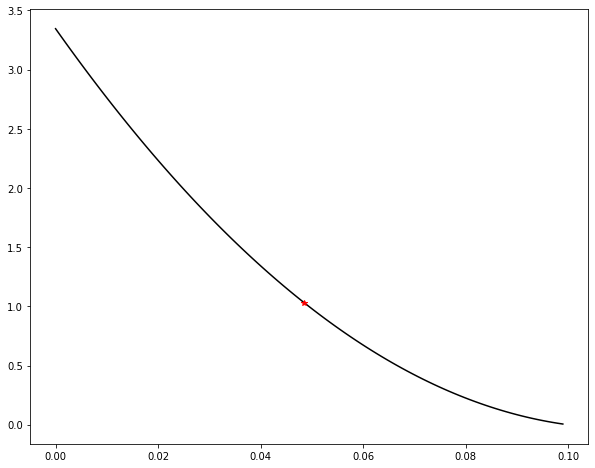}
\hspace{.5cm}
\includegraphics[width=.45\linewidth]{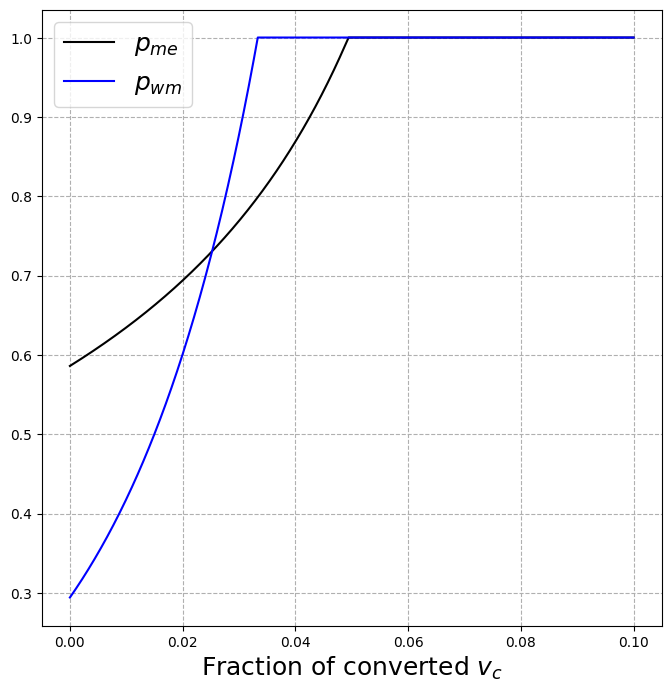}
\caption{Left: $R_{*,e}$ as a function of $v_c$. The red asterisk is the intersection point with $y=1$.\\
Right: Plot of $p_{me}$ and $p_{wm}$ as a function of $v_c$. Note that this figure may be deduced from Figure \ref{C}  with $v$ restricted to $[0.9,1]$. 
This is because we fix the initial fraction of vaccinated $v_i$ and then  define $v=v_i+v_c$ and compute $p_{me}$ and $p_{wm}$ with such combined $v$.
}
\label{converted}
\end{figure}

Up to now, we have assumed  that the introduction of mandates in a society will not change the vaccine percentage. This is not realistic, since some parents might vaccinate their children in order to let them go to the school of their preference. We call those parents ``converted''. We may think of $v$ as the sum of two components: $v_i$ that is the initial proportion of vaccinated individuals and $v_c$ that is the additional portion of ``converted''. Note that $v_c \leq 1-v_i$. The graph in Figure \ref{converted} shows a plot of $R_{*,e}$ as a function of $v_c$, when $\pi=0.5$, $v_i=0.9$ and $n_c=25$.

\section{Discussion}
In this paper we study how vaccine mandates might affect the spread of an infectious disease in elementary schools, and especially how the probability of being infected changes for medically exempt children. This effect may be counter-intuitive and local mandates might worsen the risks for medically exempt children. This work was inspired by a law that was proposed by the Dutch House of Representatives and was later rejected by the Senate \cite{Eerstekamer}. In this measure, childcare owners could freely decide whether or not joining the public vaccination program was compulsory for children to enrol at the daycare. 

In our study we do not distinguish between child-daycare and elementary schools, and we only allow the disease to spread within schools, while the population structure and contact patterns are actually much more complex and the infection could take place in households, sport centers and so on. Moreover, we do not consider multiple waves of infection nor the possibility of becoming susceptible after being recovered, which indeed is not that relevant for measles, but might be for other diseases.

Despite the limitations due to the model hypotheses, the analysis describes the qualitative different behaviours of the two situations (with and without vaccine mandates) and can be important for further ethical and political decisions made on vaccines. 

The analysis can easily be extended to diseases with lower reproduction number than measles, see Appendix \ref{appendix:lower R0}. It depends on the relative spread within and between households, whether this will lead to less or more pronounced possible undesirable effects of vaccine mandates. 

\appendix
\section{Random class sizes}
\label{appendix:random class sizes}

Our analysis is easily extended to random class sizes, in which the number of children in a class is distributed as a random variable $N_c$.
Call $p_k=\mathbb{P}(N_c=k)$; then the probability that a newly infected class is of size $k$ is given by the size-biased version $\Bar{N_c}$ of $N_c$ and we define $\Bar{p}_k= \mathbb{P}(\Bar{N_c}=k)$. 
To evaluate the expected number of type-$2$ individuals that a type-$1$ individual infects (which is element $(1,2)$ of matrix $A$ in (\ref{matrix})), we compute
$$
\mathbb{E}[\Bar{N_c}-1]=\sum_{k=1}^\infty (k-1) \Bar{p}_k= \sum_{k=1}^\infty \frac{k(k-1)p_k}{\mathbb{E}[N_c]}= \frac{\mathbb{E}[N_c (N_c-1)]}{\mathbb{E}[N_c]}.
$$
For example, if $N_c \sim Poiss(m)$, say with $m=25$. Then $\mathbb{E}[N_c]=m$ and, with the same procedure as in Section \ref{Sec:Analysis}, we obtain $\lambda_G \approx 5.6$.

We  denote  the random number of unvaccinated per class after both vaccines and mandates are introduced in the model again by $X$. Then $X$ conditioned on $N_c=k$ is $Bin (k,u)$ distributed and
$$
\mathbb{P}(X=l)=\sum_k \mathbb{P}(X=l | N_c=k) \mathbb{P}(N_c=k)
$$
while
\begin{equation}\label{Xbar}
\mathbb{P}(\Bar{X}=l)=\frac{l\mathbb{P}(X=l)}{\mathbb{E}[X]}= \sum_k \mathbb{P}(\Bar{X}=l|N_c=k) \mathbb{P}(N_c=k).
\end{equation}
We observe that 
$$
\mathbb{P}(\Bar{X}=l|N_c=k)= \frac{l \mathbb{P}(X=l|N_c=k)}{\mathbb{E}[X|N_c=k]}
$$
which implies, after some computations, that $\Bar{X}$ conditioned on $N_c=k $ is distributed as $1+ Bin(k-1,u)$, and then allows us to compute the unconditioned distribution of $\bar{X}$ using \eqref{Xbar}.

\section{Non trivial local infection probabilities}
\label{appendix:pL<1}

Throughout all our analysis above, we have considered the case in which the local infection probability $p_L=1$, which simplifies the model as explained in Section \ref{Sec:Model}. Here we want to study non trivial cases with $p_L<1$, that correspond to a different contact pattern. 

Assume that an infectious individual infects their classmates with probability $p_L=1/2$ per classmate during their infectious period. Setting both $v$ and $\pi$ to zero, we compute the global infection rate $\lambda_G$ as in Section \ref{Sec:Analysis}.
We consider a class of size $25$ in which one student gets infected from a global contact. The probability that a given one of their classmates avoids infection within two steps is $\frac{1}{2}\Big(1-\frac{1}{4} \Big)^{23}$ and the probability that at least one of the classmates avoids infection  is at least $24 \times \frac{1}{2}\Big(1-\frac{1}{4} \Big)^{23} \approx 0$. This means that all the pupils of that class will have contracted the virus within two generations of infection with very high probability.
A primary case infects on average $(n_c-1)/2$ classmates which will, in turn, infect as many remaining susceptibles. The next generation matrix (\ref{matrix}) then becomes
$$
A=\begin{bmatrix}
    \lambda_G & \frac{n_c-1}{2} & 0 \\
    \lambda_G & 0 & 1 \\
    \lambda_G & 0 & 0 \\
\end{bmatrix}
$$
and $R_0$ solves
$$
R_0^2(\lambda_G - R_0) + R_0 \lambda_G \frac{n_c-1}{2}+ \lambda_G \frac{n_c -1}{2}= 0.
$$
If $R_0=15$ and $n_c=25$ then
$$
\lambda_G= \frac{R_0^3}{R_0^2 + \frac{n_c -1}{2} + R_0 \frac{n_c-1}{2}} \approx 8.1.
$$
Recall that if $p_L=1$, then $\lambda_G \approx 5.8$.

Now introduce both vaccines and mandates in the model. We consider a class with $n+1$ unvaccinated children, one of which is infected through a global contact. The epidemic that starts spreading via local contacts in the class is described by -- to use the notation of Andersson and Britton \cite[Chapter 2]{andersson2012stochastic} -- the \textit{standard SIR epidemic model} $E_{n,1}(\lambda_L,1)$, where $\lambda_L$ indicates the local infectious rate. To compute the value of $\lambda_L$ we link it to $p_L$ by observing that an infectious individual makes contacts with a given classmate at the time steps of a Poisson$\big(\lambda_L/n \big)$ process; therefore, the probability $1-p_L$ of having no contacts with a given individual is given by $e^{-\lambda_L/n}$.

\begin{figure}
\includegraphics[width=6cm]{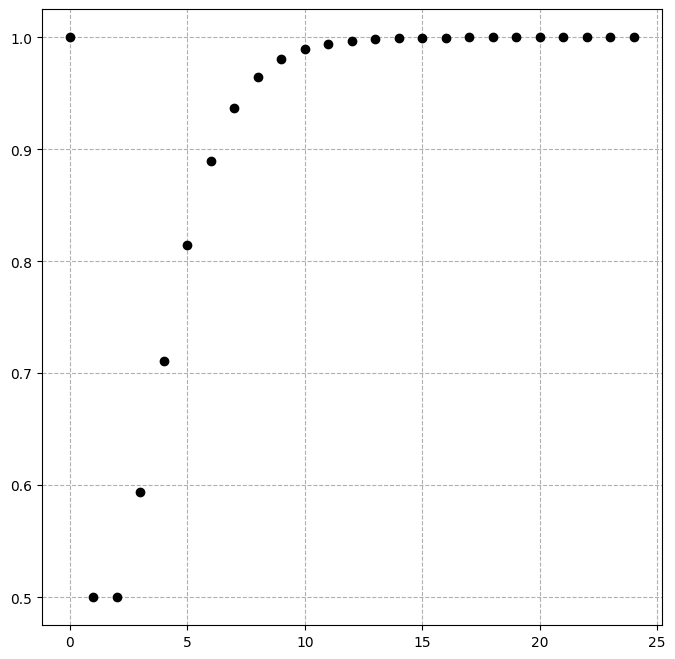}
\centering
\caption{Value of $P^n_n$ plotted as a function of $n$.}
\label{probabilities}
\end{figure}

We indicate by $P^n_j$ the probability that $j$ of the $n$ initial susceptibles get infected during the (local) epidemic. The distribution of these probabilities is obtained by solving a lower triangular linear system given in \cite[Chapter 2, Theorem 2.2]{andersson2012stochastic}. Then, in this case, the probability of extinction $\Bar{q}$ in (\ref{qbar}) will be
\begin{equation}
\begin{aligned}
\Bar{q} & =\sum_{k = 1}^{n_c} \mathbb{P}(\Bar{X}=k) \sum_{j=0}^{k-1} P^{k-1}_j \sum_{l=0}^\infty \frac{[\lambda_G(1-v) (j+1)]^l}{l!}e^{-\lambda_G(1-v) (j+1)}\Bar{q}^l\\
&=\sum_{k = 1}^{n_c} \mathbb{P}(\Bar{X}=k) \sum_{j=0}^{k-1} P^{k-1}_j e^{-\lambda_G(1-v) (j+1)(1-\Bar{q})}.
\end{aligned}
\label{Barq with pL=1/2}
\end{equation}
We observe that, keeping $p_L$ fixed and increasing the number of susceptibles (of the local epidemic) $n$, the probability $P^n_n$ that everyone gets infected gets closer and closer to $1$, as seen in Figure \ref{probabilities}.

\begin{figure}
\includegraphics[width=.45\linewidth]{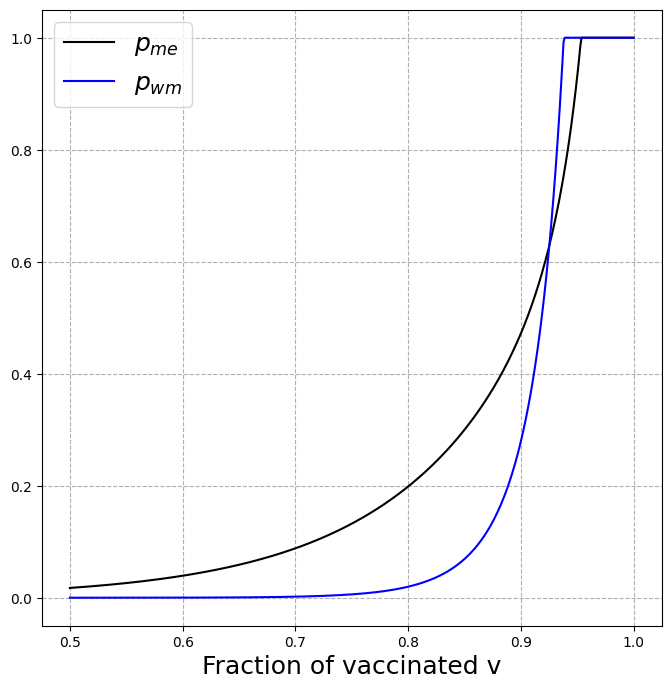}
\hspace{1cm}
\includegraphics[width=.45\linewidth]{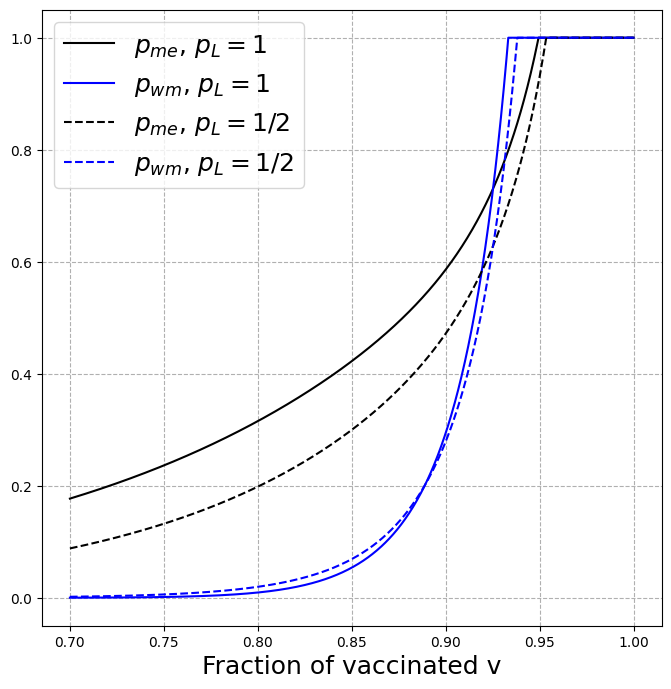}
 \caption{Left: Plot of the probability $p_{me}$ that a medically exempt child avoids infection as a function of $v$ with $\pi$ set to $0.5$, and of the probability $p_{wm}$ that a random unvaccinated child avoids infection in the case with no mandates ($\pi=0$), when the local probability $p_L$ is set to $1/2$. The two lines cross at $v_s \approx 0.925$, as in Figure \ref{C}.}

 Right: Confront between the cases $p_L=1$ (see Figure \ref{C}) and $p_L=1/2$.
 \label{pL=1/2}
\end{figure}

In Figure \ref{pL=1/2} we see, on the left, a plot of the probabilities $p_{me}$ and $p_{wm}$ that a medically exempt child avoids infection in the case with and without class mandates respectively, when $\pi=1/2$, as functions of $v$. On the right, we can see a comparison between the cases $p_L=1$ and $p_L=1/2$. We observe that, even if in the latter case the local contact probability is halved, the probability $p_{me}$ that a medically exempt child avoids infection is higher in the first case. This depends on the value of the global infection rate $\lambda_G$, which we remember increases from 5.8 in the first case to 8.1 in the second: if $R_0$ stays the same but $p_L$ decreases, then $p_G$ increases, which means that the medically exempts are more exposed to the virus. For the probability $p_{wm}$ that a random unvaccinated avoids infection in the case without mandates we conversely see almost no difference in the two cases. \newline

\begin{figure}
\includegraphics[width=.45\linewidth]{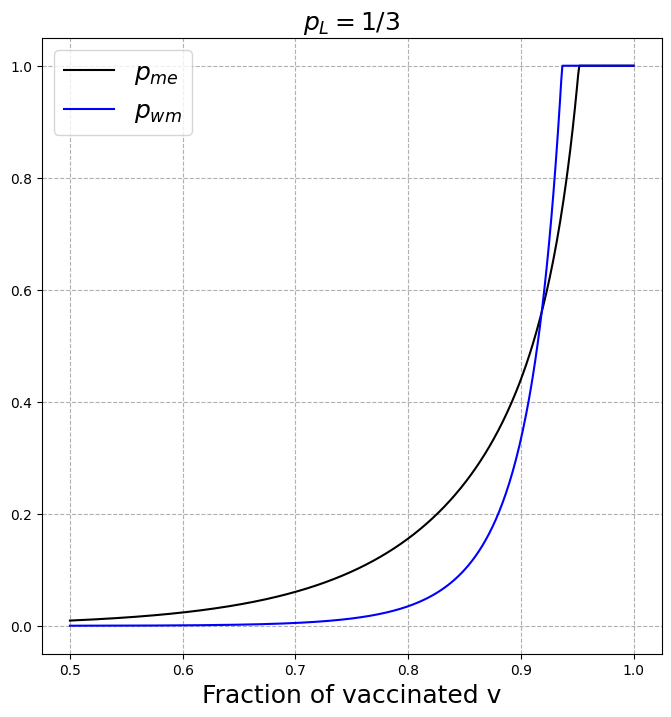}
\hspace{1cm}
\includegraphics[width=.45\linewidth]{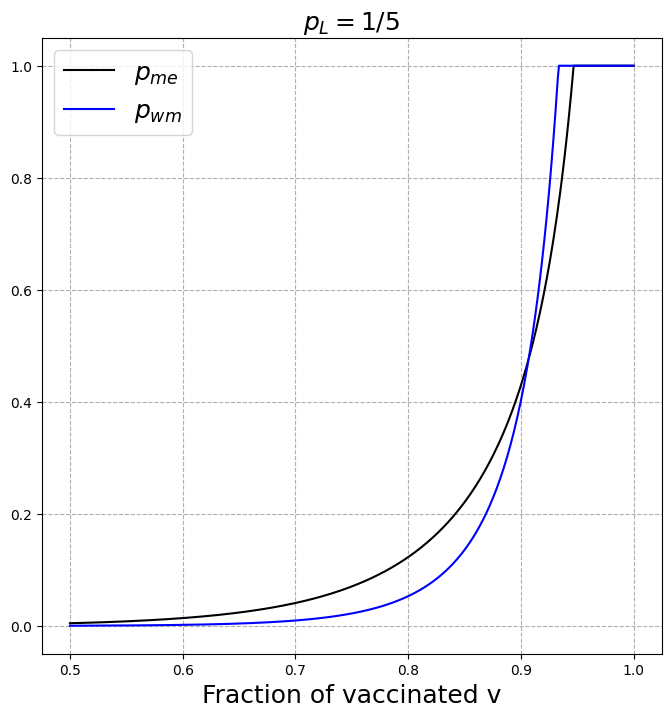}
 \caption{Plot of the probability $p_{me}$ that a medically exempt child avoids infection as a function of $v$ with $\pi$ set to $0.5$, and of the probability $p_{wm}$ that a random unvaccinated child avoids infection in the case with no mandates ($\pi=0$), when the local probability $p_L$ is set to $1/2$ (left) and $1/5$ (right).}
 \label{pL=1/3 and 1/5}
\end{figure}

Similarly we can analyse the cases $p_L=1/3$ and $p_L=1/5$. For the same reason as above, we can assume each individual in a class to be infected within a maximum of two generations of infectives. If one child is infected by a global contact, it will infect on average $(n_c-1)/3$ and $(n_c-1)/5$ classmates respectively, which will in turn infect the remaining $2(n_c-1)/3$ and $4(n_c-1)/5$ children. The next generation matrices become
$$
A_{1/3}=\begin{bmatrix}
    \lambda_G & \frac{n_c-1}{3} & 0 \\
    \lambda_G & 0 & 2 \\
    \lambda_G & 0 & 0 \\
\end{bmatrix}
\qquad
A_{1/5}=\begin{bmatrix}
    \lambda_G & \frac{n_c-1}{5} & 0 \\
    \lambda_G & 0 & 4 \\
    \lambda_G & 0 & 0 \\
\end{bmatrix}
$$
which lead to $\lambda_G \approx 9.3$ and 10.7 respectively. Then, using the same expression for $\Bar{q}$ found in (\ref{Barq with pL=1/2}), we obtain similar graphs for $p_{me}$ and $p_{wm}$, as seen in Figure \ref{pL=1/3 and 1/5}.

\begin{figure}
\includegraphics[width=.45\linewidth]{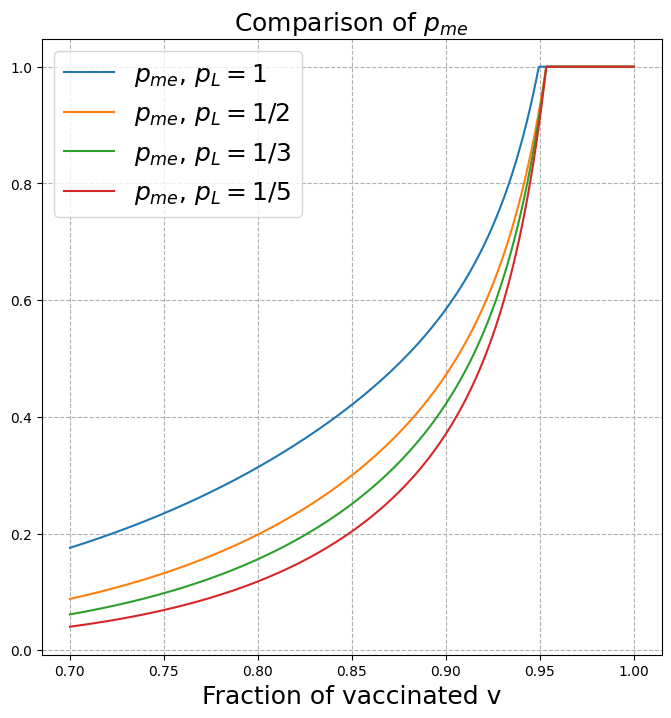}
\hspace{1cm}
\includegraphics[width=.45\linewidth]{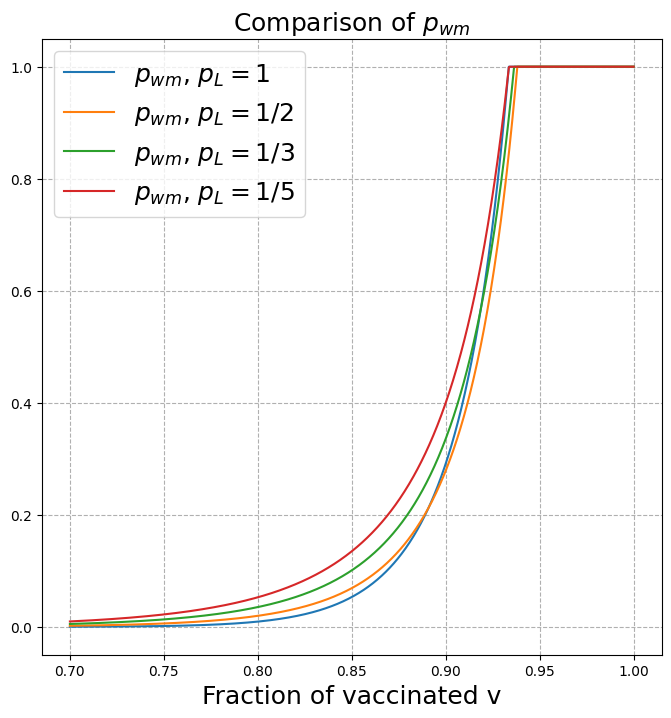}
 \caption{Comparison of the plots of the probability $p_{me}$ that a medically exempt child avoids infection as a function of $v$ when $\pi=0.5$ (left), and of the probability $p_{wm}$ that a random unvaccinated child avoids infection in the case of no vaccine mandates (right), for the different $p_L$ values considered above.}
 \label{comparison}
\end{figure}

Finally, in Figure \ref{comparison} we can see a comparison between the plots for $p_{me}$ (on the left) and $p_{wm}$ (on the right) for the different $p_L$ values we have previously considered. What we observe is that, while $p_{me}$ tends to decrease when $p_L$ decreases, $p_{wm}$ has the opposite behaviour. When mandates are not present and unvaccinated are uniformly spread in the classrooms, having a lower $p_L$ value slightly increases the probability $p_{wm}$ of avoiding infection, as the clusters of infectives forming during the spread of the virus are smaller. When mandates are present, the probability $p_{me}$ that a medically exempt child avoids infection decreases with $p_L$, because the global infection rate $\lambda_G$ reaches higher values.

\section{Less infectious diseases}
\label{appendix:lower R0}

We are now interested in considering the case of a less infectious disease; for example, we take the basic reproduction number $R_0=6$.

\begin{figure}
\includegraphics[width=6cm]{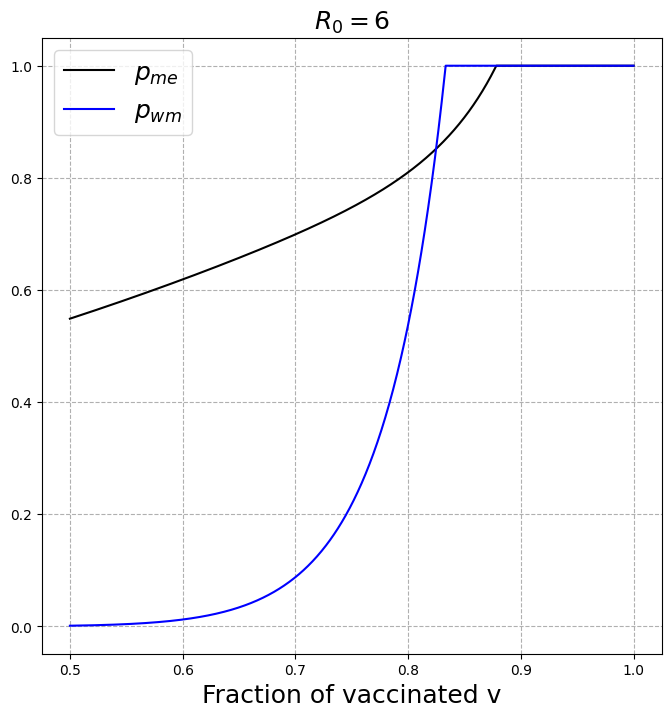}
\centering
 \caption{Plot of the probability $p_{me}$ that a medically exempt child avoids infection as a function of $v$ with $\pi$ set to $0.5$, and of the probability $p_{wm}$ that a random unvaccinated child avoids infection in the case with no mandates ($\pi=0$). We see that the two lines cross at $v_s \approx 0.82$. The minimal value of $v$ for which  $p_{wm}=1$ is at $v \approx 0.83$, while the minimal value of $v$ for which  $p_{me}=1$ is at $v \approx 0.88$. To compare it with the case $R_0=15$ see Figure \ref{C}.} 
\label{R0=6}
\end{figure}

Comparing Figure \ref{C}
and Figure \ref{R0=6} we see that in the first case the functions describing $p_{me}$ and $p_{wm}$ increase much faster than in the case $R_0=6$. This means that, with a less infectious virus, the difference between the cases with and without mandates is more pronounced. In fact, as shown in Figure \ref{R0=6}, for a medically exempt child escaping infection is significantly more probable in the case with vaccine mandates, up to the switching point $v \approx 0.82$.

\begin{figure}
\includegraphics[width=.45\linewidth]{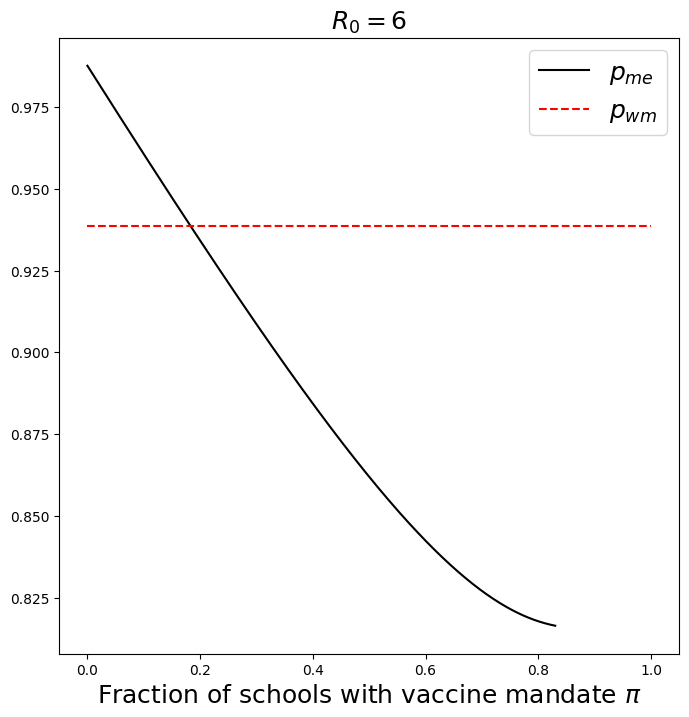} \hspace{1cm}
\includegraphics[width=.45\linewidth]{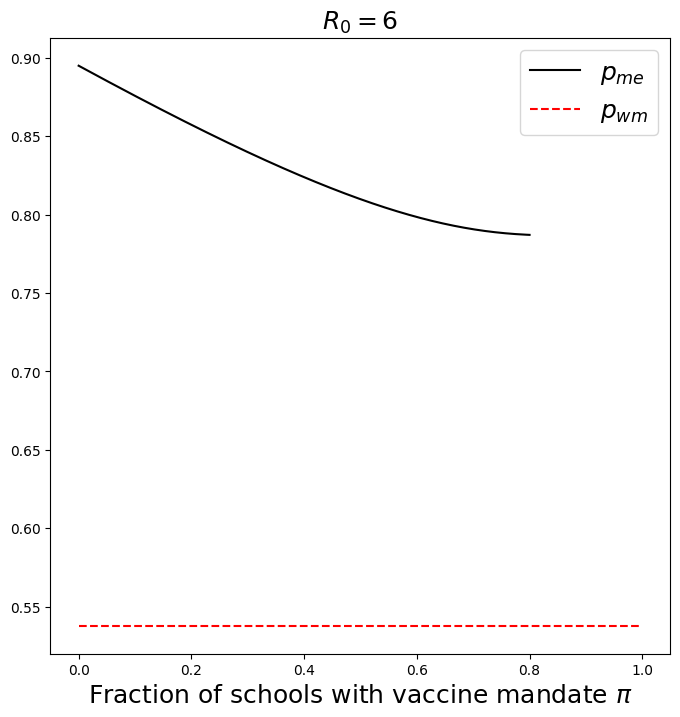}
 \caption{Plot of $p_{me}$, the probability that a medically exempt child avoids infection as a function of $\pi$, when $v=0.83$ (left) and when $v=0.80$ (right). We compare it with the probability $p_{wm}$ that a random unvaccinated child avoids infection in the case with no mandates ($\pi=0$). See Figure \ref{D} to compare it with the case $R_0=15$. }
\label{R0=6 with pi variable}
\end{figure}

In Figure \ref{R0=6 with pi variable} we see how $p_{me}$ and $p_{wm}$ vary as functions of $\pi$ when the value of $v$ is fixed and close to the herd immunity; the behaviour is the same as the one observed in Figure \ref{D}.

\newpage
\bibliographystyle{abbrv}
\bibliography{allpapers}

\end{document}